\documentclass{article}

\pdfoutput=1

\usepackage[preprint]{neurips_2023}
\usepackage[table,xcdraw]{xcolor}
\usepackage{natbib}
\setcitestyle{numbers,square}

\usepackage[utf8]{inputenc} 
\usepackage[T1]{fontenc}    
\usepackage{hyperref}       
\usepackage{url}            
\usepackage{booktabs}       
\usepackage{amsfonts}       
\usepackage{nicefrac}       
\usepackage{microtype}      
\usepackage{xcolor}         
\usepackage{amsfonts}
\usepackage{bm}
\usepackage[namelimits]{amsmath}
\usepackage{subfigure}
\usepackage[graphicx]{realboxes}
\usepackage{multirow}
\usepackage[linesnumbered,ruled,vlined]{algorithm2e}
\usepackage{enumitem}
\usepackage{comment}

\title{Diffusing on Two Levels and Optimizing for Multiple Properties: A Novel Approach to Generating Molecules with Desirable Properties}

\author{%
	Siyuan Guo\\
	Tongji University\\
	Shanghai, China \\
	\texttt{gsy990124@tongji.edu.cn} \\
	\And
	Jihong Guan* \\
	Tongji University \\
	Shanghai, China \\
	\texttt{jhguan@tongji.edu.cn } \\
        \And
        Shuigeng Zhou*\\
        Fudan University\\
        Shanghai, China\\
        \texttt{sgzhou@fudan.edu.cn }        
}

\begin{document}

	\maketitle

	\begin{abstract}
		
In the past decade, Artificial Intelligence (AI) driven drug design and discovery has been a hot research topic in the AI area, where an important branch is molecule generation by generative models, from GAN-based models and VAE-based models 
to the latest diffusion-based models. However, most existing models pursue only the basic properties like \textit{validity} and \textit{uniqueness} of the generated molecules, a few go further to explicitly optimize one single important molecular property (e.g. QED or PlogP), which makes most generated molecules little usefulness in practice. In this paper, we present a novel approach to generating molecules with desirable properties, which expands the diffusion model framework with multiple innovative designs. The novelty is two-fold. On the one hand, considering that the structures of molecules are complex and diverse, and molecular properties are usually determined by some substructures (e.g. pharmacophores), we propose to perform diffusion on two structural levels: molecules and molecular fragments respectively, with which a mixed Gaussian distribution is obtained for the reverse diffusion process. And to get desirable molecular fragments, we develop a novel \textit{electronic effect} based fragmentation method. On the other hand, we introduce two ways to explicitly optimize multiple molecular properties under the diffusion model framework. First, as potential drug molecules must be chemically valid, we optimize molecular validity by an energy-guidance function. Second, since potential drug molecules should be desirable in various properties
, we employ a multi-objective mechanism to optimize multiple molecular properties simultaneously. Extensive experiments with two benchmark datasets QM9 and ZINC250k show that the molecules generated by our proposed method have better \textit{validity, uniqueness, novelty, Fréchet ChemNet Distance (FCD), QED, and PlogP} than those generated by current SOTA models.	
\end{abstract}
	
\section{Introduction}
The fundamental goal of drug design and discovery is to find candidate molecules that have the desirable  chemical properties. Traditional drug design relies heavily on the expertise of pharmaceutical professionals. With a high failure rate up to 96\%, this process can take decades of years and cost billions of dollars~\cite{munos2011revive,paul2010improve}. With the remarkable advances in artificial intelligence (AI) techniques, the situation has been changing. In the past decade, an increasing number of AI models and algorithms have been involved in drug design and discovery, which brings in the blossoming area of AI-driven drug design and discovery. Among these efforts, a hot and promising research direction is automatic molecule generation by generative models. Especially, by efficiently exploring the vast chemical space and accurately modeling various molecules, recent deep generative models have been blazing a trail in accelerating the drug discovery process and reducing cost. Typically, these models first learn a continuous latent space by encoding the training molecules 
and then generate new molecules by decoding the learned latent space~\cite{weininger1989smiles,zang2020moflow,gomez2018automatic}. Up to now, a number of molecule generation models have been developed, which roughly fall into four types, including GAN-based models, VAE-based models, energy-based models, flow-based models, and the latest diffusion-based models.

However, most existing molecule generation models stop at pursuing high basic properties like \textit{validity} and \textit{uniqueness} of the generated molecules, only a few go further to explicitly optimize one single more important property such as QED or PlogP. As potential drug molecules should be desirable for all properties, so most generated molecules of current models are actually of limited usefulness. To challenge this situation, in this paper we try to generate molecules with multiple desirable properties, i.e., we generate molecules with multiple properties optimized simultaneously.

To this end, we propose a novel approach by expanding the diffusion model framework with multiple innovative designs. The novelty of our approach lies in two aspects:

On the one hand, considering that molecules are complex and diverse in structure, and molecular properties are heavily dependent on some substructures (e.g. pharmacophores) of molecules, we think that directly modeling the whole molecular space with a limited number of real molecules may be not enough. Different from the existing diffusion-based models, we perform diffusion on two structural levels: molecules and molecular fragments respectively. A mixed Gaussian distribution is obtained from the two levels' diffused Gaussian distributions and applied to the reverse diffusion process. As we hope that the fragments are really related to molecular properties, instead of using existing molecule fragmentation scheme like BRICS~\cite{degen2008art}, we develop a new fragmentation method based on \textit{electronic effect}, which is a physicochemical measure of molecules or molecular fragments, related to their acidity and alkalinity.

On the other hand, we design two strategies to optimize multiple properties simultaneously. First, considering that potential drug molecules must be chemically valid, we optimize the validity of molecules by an energy-guidance function to control the denoising process. Second, as potential drug molecules should be desirable in various properties, 
routine strategies that combine different properties into one indicator to be optimized, do not work well. Instead, 
we employ a multiple-objective mechanism to optimize multiple molecular properties simultaneously so that all properties can reach a desirable state.

To demonstrate the effectiveness and advantages of the proposed method, we carry out extensive experiments with two benchmark datasets QM9~\cite{ramakrishnan2014quantum} and ZINC250k~\cite{irwin2012zinc}. Seven typical existing models, covering SOTA VAE-based model~\cite{jin2018junction}, flow-based models~ \cite{madhawa2019graphnvp,shi2020graphaf,luo2021graphdf,zang2020moflow}, and diffusion-based models~\cite{jo2022score,vignac2022digress} are used for performance comparison. 
Comprehensive ablation studies are also conducted to validate the effects of different model designs and parameter settings.

Our contributions are summarized as follows: 
%
  1) We propose a novel approach to generating molecules with desirable properties. To the best of our knowledge, this is the first work that optimizes multiple properties of the generated molecules simultaneously.
  2) To model the molecules better, we develop a generative model that diffuses on two molecular structural levels--- molecules and molecular fragments respectively. We also introduce a new molecule fragmentation method based on electronic effect.
  3) We employ two ways to optimize multiple properties of generated molecules under the diffusion model framework. On the one hand, we use an energy guidance function to optimize molecular validity. On the other hand, we adopt a multiple-objective strategy to optimize other properties.
  4) We conduct extensive experiments with two benchmark datasets. The results show that compared to typical existing models, the proposed method can generate molecules with better validity, uniqueness, novelty, Fréchet ChemNet Distance (FCD), QED, and PlogP.

\section{Related Work}
Existing deep generative models for molecule generation fall roughly into five types, including GAN-based, VAE-based, energy-based, flow-based, and recent diffusion-based models.

\textbf{GAN-based models.}
Early works of deep molecular generation models are mostly based on GANs, where the distribution of the training set is implicitly learned by the competition learning of the generator and the discriminator. Typical GAN-based models include ORGAN~\cite{guimaraes2017objective}, MolGAN~\cite{de2018molgan}, and GA-GAN~\cite{blanchard2021using}. These models can achieve high novelty due to implicit and likelihood-free features, but their validity/uniqueness is limited. Due to the inherent problems with GANs such as perfect discriminator and mode collapse, it is challenging to train GAN-based models, and difficult to generate molecules with desirable molecular properties.

\textbf{VAE-based models.}
Up to now, a number of VAE-based molecular generation models have been developed, including GraphVAE~\cite{simonovsky2018graphvae}, JTVAE~\cite{jin2018junction}, CGVAE~\cite{liu2018constrained}, and NeVAE~\cite{samanta2020nevae}. In these models, the data distribution is learned by the identification model (encoder) and the generation model (decoder)~\cite{kingma2013auto}. Although the optimization of VAE-based models is easier and more stable than GAN-based models~\cite{de2018molgan}, only the lower bound of log-likelihood is optimized, which thus limits the generative power of the networks.

\textbf{Energy-based models.}
These models try to model the energy function of molecules, by assigning lower energies to data points corresponding to real molecules and higher energies to other data points. GraphEBM~\cite{liu2021graphebm} is a representative of such models. Energy-based models are usually difficult to train due to the slow sampling process. To optimize a energy model, many estimations must be done, the inefficiency problem of energy model training will be exacerbated~\cite{du2019implicit}.

\textbf{Flow-based models.}
They directly optimize the log-likelihood function to model the generative probability of samples. Typical flow-based molecular generation models are GraphAF~\cite{shi2020graphaf}, GraphDF~\cite{luo2021graphdf} and MoFlow~\cite{zang2020moflow}. Most of these models fail to capture the permutation-invariance property of molecular graphs~\cite{liu2021graphebm} and are computationally expensive.

\textbf{Diffusion-based models.} 
They stand for the cutting-edge deep generative models, which gradually perturb the training dataset through a forward diffusion process, and then sample new data from the same distribution by gradually denoising in the reverse diffusion process. With the advantages of tractable training, flexible expansion, and permutation-invariant, such models have excellent performance in molecule generation. GDSS~\cite{jo2022score} captures the joint distribution of molecular nodes and edges, only to generate mere molecular imitations close to the training distribution. DiGress~\cite{vignac2022digress} is more efficient than GDSS by a discrete diffusion process, but the properties of generated molecules are relatively low.

\textbf{Differences between existing models and our work.} First, different from existing models, we try to generate molecules with multiple desirable properties, not only basic properties like validity and uniqueness but also more important properties such as QED and PlogP. Second, different from current diffusion-based models, we perform diffusion on two structural levels --- molecules and fragments respectively, and use the resulting mixed Gaussian distribution for the reverse process. Third, our work is the first that uses multiple-objective optimization to coordinate multiple properties.

\section{Methodology}
In this section, we first describe the framework of our work to give an overview of the proposed method. Then, we introduce our techniques in detail. 

\subsection{Overview}
\begin{figure}[!h]
 \centering
 \includegraphics[width=1\linewidth]{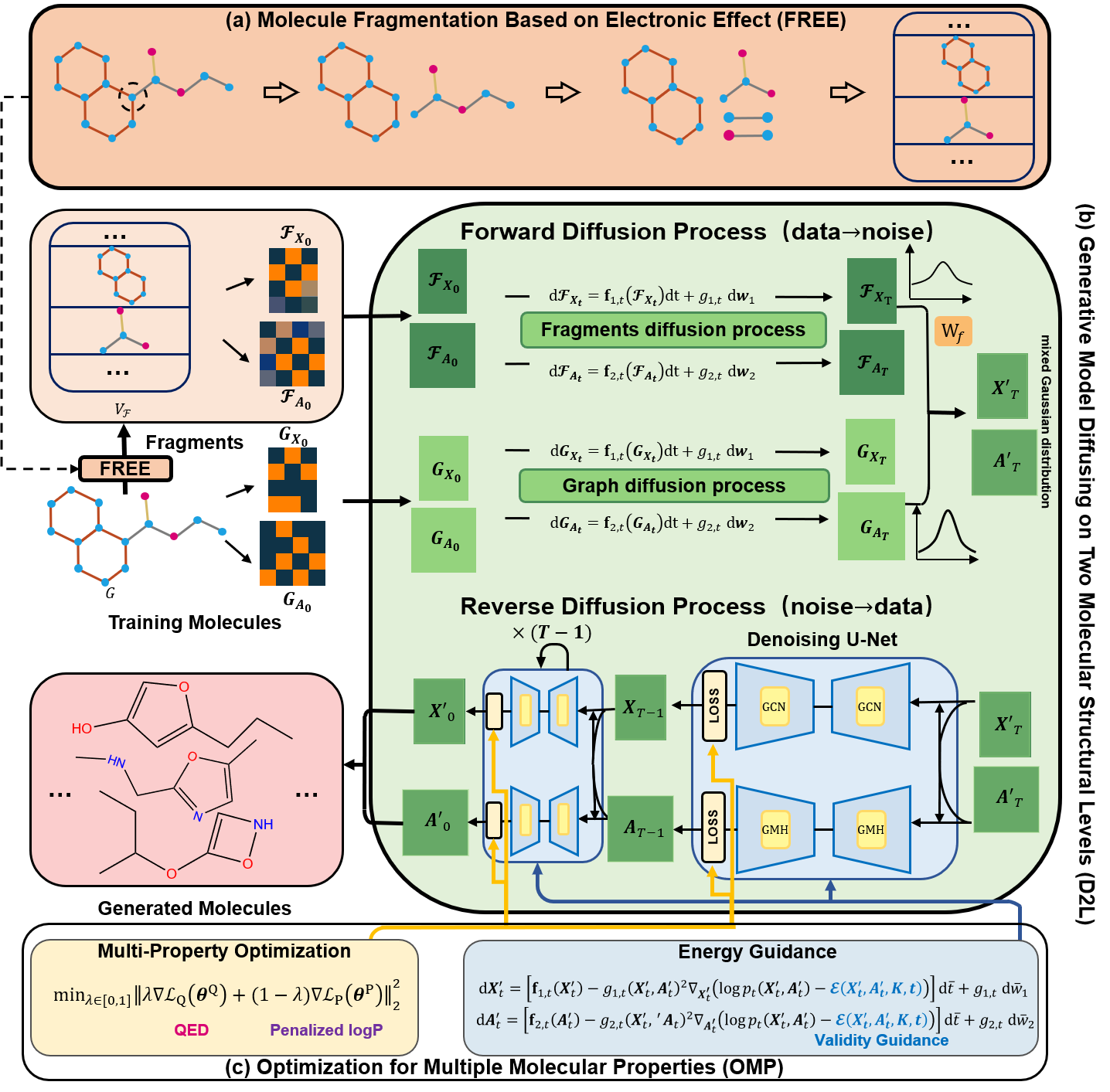}\caption{The framework of the D2L-OMP method.}\label{fig:framework}
\end{figure}

Fig.~\ref{fig:framework} shows the framework of our D2L-OMP method, which is an expanded diffusion model with multiple innovative designs. It consists of three major modules: (a) the method of molecule \textit{fragmentation based on electronic effect} (\textbf{FREE} in short), (b) the \textit{generative model diffusing on two molecular structural levels} (\textbf{D2L} in short), and (c) the module of \textit{optimization for multiple molecular properties} (\textbf{OMP} in short). Each training molecule is fragmented by the FREE method, which will be introduced in Sec.~\ref{sec:free}. All the fragments generated by FREE form a fragment vocabulary used for model training. Take all fragments in the vocabulary and the training molecules as input, the model is trained by first diffusing on two molecular structural levels --- the fragments and the molecules respectively, from which two Gaussian distributions are obtained, which are further combined to get a mixed Gaussian distribution, then denoising the mixed Gaussian distribution in the reverse diffusion process. The model will be detailed in Sec.~\ref{sec:d2l}. During model training, we try to optimize multiple molecular properties simultaneously by two strategies, which are presented in Sec.~\ref{sec:energy-guidance} and Sec.~\ref{sec:moo} respectively. On the one hand, we optimize the validity of generated molecules based on an energy guidance function, as validity is a basic property of generated molecules. On the other hand, we employ multi-objective optimization to optimize multiple molecular properties, which can make multiple properties desirable. 
In this paper, without loss of generality, we optimize validity, QED, and PlogP. Actually, our method can be expanded to optimize more properties.    

	
\subsection{Molecule Fragmentation Based on Electronic Effect}\label{sec:free}
Generally, we hope that the fragments are closely associated with molecular properties. Actually, there are already a number of molecule fragmentation schemes in the literature, including BRICS~\cite{degen2008art}. However, our experiments show that fragments generated by those schemes are not satisfactory in our work (also see the ablation results in Sec.~\ref{sec:ablation}). So we propose a new fragmentation method based on \textit{electronic effect}~\cite{miessler2014inorganic}. 

Electronic effects happen when the substituents of molecules cause the distribution change of the electron cloud density within molecules, consequently impacting the acidity or alkalinity of substituents and molecules. Substituents are qualified by the substituent constant. The Hammett equation, which describes the relationship between the substituent constant and the equilibrium constant when the reaction type is determined, is used to calculate the substituent constant $\sigma$~ \cite{hammett1937effect,iupac1997compendium,keenan2008determination}:
    $\sigma=\frac{1}{\rho}log\dfrac{K}{K_0},$
where $\rho$ is the reaction constant, which is related only to the type of reaction. $K$ is the equilibrium constant when the substituent is not a hydrogen atom, and $K_0$ is the equilibrium constant when the substituent is a hydrogen atom. In~\cite{iupac1997compendium}, various substituents and their constants are presented in the so-called Hammett table. 

Given a molecule, we fragment it as follows: 1) Split all ring substructures from their chain substructures, resulting in a set $R$=\{$R_i$\} of ring substructures and a set $C$=\{$C_i$\} of chain substructures, $R_i$ and $C_i$ are a ring substructure and a chain substructure respectively. 2) We directly accept all elements in $R$ as fragments. 3) For each element $C_i$ in $C$, first use the substituent with the highest constant in the Hammett table to search matching substructures in $C_i$, and accept the hits as fragments, meanwhile removing the matched substructures from $C_i$, and denoting the remaining as $C_i'$; Then, use the substituent with the second highest constant in the Hammett table to search matching substructures in $C_i'$, and save the hits as fragments. The above process goes iteratively till no more match can be found. 4) Using $\sigma_R$ as a threshold to filter out the fragments with constant $\left|\sigma\right|$ $<$ $\sigma_R$, the remaining fragments form the final fragment set. In our experiments, we set the $R=$hydrogen atom, i.e., using the $\sigma_H$ as the threshold value. In such a way, we can get all fragments from the training molecules, which form a \textit{fragment vocabulary} $V_{\boldsymbol{\mathcal{F}}}$. In this paper, we call $V_{\boldsymbol{\mathcal{F}}}$ the \textit{FREE vocabulary}, and all fragments \textit{FREE fragments}.

\subsection{Diffusion on Two Molecular Structural Levels}\label{sec:d2l}
For a molecular graph $\boldsymbol{G}$ with $m$ nodes (atoms), we have $\boldsymbol{G}=(\boldsymbol{G}_X,\boldsymbol{G}_A)$ where $\boldsymbol{G}_X\in{\mathbb{R}^{m\times{dim}}}$ and $\boldsymbol{G}_A\in {\mathbb{R}^{m\times{m}}}$ are the node feature matrix and the weighted adjacency matrix respectively, $dim$ is the dimension of the node features. Similarly, for a FREE fragment $\boldsymbol{\mathcal{F}}$ with $n$ nodes, we have $\boldsymbol{\mathcal{F}}=(\boldsymbol{\mathcal{F}}_X,\boldsymbol{\mathcal{F}}_A)$ where $\boldsymbol{\mathcal{F}}_X\in{\mathbb{R}^{n\times{dim}}}$ and $\boldsymbol{\mathcal{F}}_A\in {\mathbb{R}^{n\times{n}}}$ are the node feature matrix and the weighted adjacency matrix respectively. To train the D2L-OMP model, all training molecular graphs and the corresponding FREE fragments are used, on which the diffusion is performed respectively.  
%

By step-wisely injecting noise, the training molecular graphs and fragments are respectively and smoothly transformed into a Gaussian distribution, they are then combined into a mixed Gaussian distribution. Inspired by~\cite{jo2022score}, to successfully generate samples from the data distribution, the node feature matrix and weighted adjacency matrix of each molecular graph should be diffused simultaneously. Denote the forward diffusion process on the molecular graph $\boldsymbol{G}$ as $\{\boldsymbol{G}_t=(\boldsymbol{G}_{X_t},\boldsymbol{G}_{A_t})\}_{t=0}^T$ in a fixed time horizon $t\in[0,T]$, which can be modeled as the solution to an $It$$\widehat{o}$ SDE:
	\begin{equation}
		\mathrm{d}\boldsymbol{G}_t=\mathbf{f}_t(\boldsymbol{G}_t)\mathrm{d}t+ g(t)\mathrm{d}\boldsymbol{\mathrm{w}},\boldsymbol{G}_0\sim p_{mol},
		\label{equ:equ2}
	\end{equation}
	where $\boldsymbol{G}_0$ is a graph from the data distribution $p_{mol}$, $\mathbf{f}_t(\cdot)$, $g(\cdot)$ and $\boldsymbol{\mathrm{w}}$ are the linear drift coefficient, the diffusion coefficient, and the standard Wiener process, respectively. In Eq.~(\ref{equ:equ2}), at each infinitesimal time step $\mathrm{d}t$, an infinitesimal noise $\mathrm {d}\boldsymbol{\mathrm{w}}$ is added to $\boldsymbol{G}_{X_0}$ and $\boldsymbol{G}_{A_0}$. The coefficients $\mathbf{f}_t(\cdot)$ and $g(\cdot)$ are related to the size of noise. To efficiently generate samples, the sample $\boldsymbol{G}_t$ approximately follows a prior distribution with a tractable form (e.g. a Gaussian distribution).
	
Similarly, the forward diffusion process on a FREE fragment $\boldsymbol{\mathcal{F}}$ can be denoted as $\{\boldsymbol{\mathcal{F}}_t=(\boldsymbol{\mathcal{F}}_{X_t},\boldsymbol{\mathcal{F}}_{A_t})\}_{t=0}^T$ in a fixed time horizon $t\in[0,T]$, which can be modeled as the solution to an $It$$\widehat{o}$ SDE:
	$\mathrm{d}\boldsymbol{\mathcal{F}}_t=\mathbf{f}_t(\boldsymbol{\boldsymbol{\mathcal{F}}}_t)\mathrm{d}t+g(t)\mathrm{d}\boldsymbol{\mathrm{w}},\boldsymbol{\mathcal{F}}_0\sim p_\mathcal{F}$.
	
Since the sum of two independent Gaussian distributions is still a Gaussian distribution.We expand the dimension of  $\boldsymbol{\mathcal{F}}_T$ to be consistent with the $\boldsymbol{G}_T$. The diffused sample $\boldsymbol{G}_T=(\boldsymbol{G}_{X_T},\boldsymbol{G}_{A_T})$ and $\boldsymbol{\mathcal{F}}_T=(\boldsymbol{\mathcal{F}}_{X_T},\boldsymbol{\mathcal{F}}_{A_T})$ are mixed into a new Gaussian distribution $\boldsymbol{G'}_T=(\boldsymbol{X'}_T,\boldsymbol{A'}_T)$:
	\begin{equation}
		\boldsymbol{G'}_T=\boldsymbol{G}_T+\mathrm{W}_f\boldsymbol{\mathcal{F}}_T,
	\end{equation}
	\begin{equation}
		\left\{\begin{array}{c}
			\boldsymbol{X'}_T=\boldsymbol{G}_{X_T}+\mathrm{W}_f\boldsymbol{\mathcal{F}}_{X_T},\\
			\boldsymbol{A'}_T=\boldsymbol{G}_{A_T}+\mathrm{W}_f\boldsymbol{\mathcal{F}}_{A_T},
		\end{array}\right.
	\end{equation}
	where $\mathrm{W}_f$ is a weight (or hyperparameter) to control the influence of FREE fragments on the mixed Gaussian distribution. 
	
\subsection{Energy Guidance Function}\label{sec:energy-guidance}
To promote the generation of molecules of high validity, in this section, we introduce an energy guidance function as a constraint on the generation process where the noise added in the forward diffusion process is step-wisely removed during the reverse diffusion process. Because the energy function can capture dependencies between two variables~\cite{lecun2006tutorial}, we use low energy to indicate that the generated molecules meet the desired properties (e.g. high validity), which in turn encourages the model to generate more low-energy molecules while discarding the high-energy ones, thus generating more valid chemical molecules.

As potential drug molecules must be chemically valid, we design the energy guidance function as
    \begin{equation}
    \mathcal{E}(\boldsymbol{X},\boldsymbol{A},K,t)=C_V\mathcal{E_V}(\boldsymbol{X'}_t,\boldsymbol{A'}_t,K_V,t),
    \end{equation}
where $K_V$ represents desired validity. $C_V$ is a hyper-parameter to control the optimization strength of $K_V$. $\mathcal{E_V}(\cdot,\cdot,\cdot,\cdot)$ is the distance function, measuring the gap between the predicted property and $K_V$. 

Set $\mathcal{E_V}(\cdot,\cdot,\cdot,\cdot)$ as the squared error between the predicted property and $K_V$:
   \begin{equation}
    \mathcal{E}(\boldsymbol{X},\boldsymbol{A},K,t)=C_V\left|g_v\left(\boldsymbol{X'}_t,\boldsymbol{A'}_t, t\right)-K_V\right|^2,
    \label{equ:equ6}
    \end{equation}
where $g_v(\cdot,\cdot,\cdot)$ is a discriminant function of validity, which requires additional chemistry domain knowledge to construct. In our experiments, we use the sanitizeMol package in RDKit to implement it.  Eq.~(\ref{equ:equ6}) can assign low energy to the molecules with high validity $K_V$ while assigning high energy to those of low validity. Here, although we focus on validity, the energy guidance function can include other chemical properties, if needed. For convenience, we denote 
    $\mathcal{E}(\boldsymbol{X'}_t,\boldsymbol{A'}_t,K,t)$ by $\mathcal{E}_{val}$.

In order to apply the above energy guidance function $\mathcal{E}_{val}$ to guide the reverse diffusion process, let $p_t(\cdot)$ be the marginal distribution under the forward diffusion process at time $t$, which can be written as the following equation: 
	\begin{equation}
		\left\{\begin{array}{c}
			\mathrm{d} \boldsymbol{X'}_t =\left[\mathbf{f}_{1, t}\left(\boldsymbol{X'}_t\right)
			-g_{1, t}(\boldsymbol{X'}_t, \boldsymbol{A'}_t)^2 \nabla_{\boldsymbol{X'}_t} (\log p_t\left(\boldsymbol{X'}_t, \boldsymbol{A'}_t\right)-\mathcal{E}_{val})\right] \mathrm{d} \bar{t}+g_{1, t} \mathrm{~d} \bar{\boldsymbol{\mathrm{w}}}_1 \\
			\mathrm{d} \boldsymbol{A'}_t =\left[\mathbf{f}_{2, t}\left(\boldsymbol{A'}_t\right)
			-g_{2, t}(\boldsymbol{X'}_t, \boldsymbol{'A}_t)^2 \nabla_{\boldsymbol{A'}_t} (\log p_t\left(\boldsymbol{X'}_t, \boldsymbol{A'}_t\right)-\mathcal{E}_{val})\right] \mathrm{d} \bar{t}+g_{2, t} \mathrm{~d} \bar{\boldsymbol{\mathrm{w}}}_2
		\end{array}\right.
	\label{equ:equ8}
	\end{equation}
where $\mathbf{f}_{1, t}(\cdot)$ and $\mathbf{f}_{2, t}(\cdot)$ satisfy $\mathbf{f}_t(\boldsymbol{G'})=\mathbf{f}_t(\boldsymbol{X'},\boldsymbol{A'}) =(\mathbf{f}_{1, t}\left(\boldsymbol{X'}_t\right),\mathbf{f}_{2, t}\left(\boldsymbol{A'}_t\right))$. $g_{1, t}$ and $g_{2, t}$ are scalar diffusion coefficients. $\bar{\boldsymbol{\mathrm{w}}}_1$ and $\bar{\boldsymbol{\mathrm{w}}}_2$ are the inverse time standard Wiener process. Eq.~(\ref{equ:equ8}) describes the simultaneous diffusion processes of the node feature matrix $\boldsymbol{X'}$ and the adjacency matrix $\boldsymbol{A'}$ that are associated with time. $\nabla_{\boldsymbol{X'}_t} \log p_t\left(\boldsymbol{X'}_t, \boldsymbol{A'}_t\right)$ and $\nabla_{\boldsymbol{A'}_t} \log p_t\left(\boldsymbol{X'}_t, \boldsymbol{A'}_t\right)$ are the marginal distribution scores of $\boldsymbol{X'}$ and $\boldsymbol{A'}$, respectively. When these two scores are known for all $t$, the simulation of Eq.~(\ref{equ:equ8}) starts to sample from $p_0$.

The scores $\nabla_{\boldsymbol{X'}_t} \log p_t\left(\boldsymbol{X'}_t, \boldsymbol{A'}_t\right)$ and $\nabla_{\boldsymbol{A'}_t} \log p_t\left(\boldsymbol{X'}_t, \boldsymbol{A'}_t\right)$ are estimated by training the time-dependent score-based models $\boldsymbol{s}_\theta(\cdot,t)$ and $\boldsymbol{s}_\psi(\cdot,t)$ on samples, because the ground-truth score of either marginal distribution is analytically accessible. $\boldsymbol{s}_\theta(\cdot,t)$ and $\boldsymbol{s}_\psi(\cdot,t)$ are trained to reach a minimum Euclidean distance to the ground-truth scores by applying the method of denoising score matching~\cite{vincent2011connection,song2020score}, which can be represented by the equations below:
	\begin{equation}
    	\begin{gathered}
    		\min _\theta \mathbb{E}_t\left\{\lambda_1(t) \mathbb{E}_{\boldsymbol{G}_0} \mathbb{E}_{\boldsymbol{G'}_t \mid \boldsymbol{G}_0}\left\|\boldsymbol{s}_{\theta}\left(\boldsymbol{G'}_t,t\right)-\nabla_{\boldsymbol{X'}_t} \log p_{0 t}\left(\boldsymbol{X'}_t \mid \boldsymbol{X}_0\right)\right\|_2^2\right\} \\
    		\min _\psi \mathbb{E}_t\left\{\lambda_2(t) \mathbb{E}_{\boldsymbol{G}_0} \mathbb{E}_{\boldsymbol{G'}_t \mid \boldsymbol{G}_0}\left\|\boldsymbol{s}_{\psi}\left(\boldsymbol{G'}_t,t\right)-\nabla_{\boldsymbol{A'}_t} \log p_{0 t}\left(\boldsymbol{A'}_t \mid \boldsymbol{A}_0\right)\right\|_2^2\right\} .
    	\end{gathered}
    \end{equation}
   where $\lambda_1(t)$ and $\lambda_2(t)$ are positive weighting functions, and $t$ is uniformly sampled from $[0,T]$.
   
    Then, we introduce the score-based model $\boldsymbol{s}_\theta(\cdot,t)$ to estimate $\nabla_{\boldsymbol{X'}_t} \log p_t\left(\boldsymbol{X'}_t,\boldsymbol{A'}_t\right)$, which has the same dimensionality as $\boldsymbol{X'}_t$. Multiple layers of GNNs are used in this process:
    \begin{equation}
    	\boldsymbol{s}_{\theta}(\boldsymbol{G'}_t,t)=MLP([{H_i}_{i=0}^L])
    \end{equation}
    where $\boldsymbol{H}_{i+1}=GNN(\boldsymbol{H}_i,\boldsymbol{A'}_t)$,   $\boldsymbol{H}_0=\boldsymbol{X'}_t$, and $L$ is the number of layers of the GNNs.

    Similarly, the score-based model $\boldsymbol{s}_\psi(\cdot,t)$ is used to estimate $\nabla_{\boldsymbol{A'}_t} \log p_t\left(\boldsymbol{X'}_t,\boldsymbol{A'}_t\right)$, which has the same dimensionality as $\boldsymbol{A'}_t$. The Graph Multi-Head (GMH) attention~\cite{baek2021accurate} is utilized to distinguish the important relations between different nodes, which can be formulated as follows:
    \begin{equation}
    	 \boldsymbol{s}_{\psi}(\boldsymbol{G'}_t,t)=MLP([{GMH(\boldsymbol{H}_i,\boldsymbol{A'}_t^p)}_{i=0,p=1}^{J,P}])
    \end{equation}
    where $\boldsymbol{A'}_t^p$ is a higher-order adjacency matrix, $\boldsymbol{H}_{i+1}=GNN(\boldsymbol{H}_i,\boldsymbol{A'}_t)$,  $\boldsymbol{H}_0=\boldsymbol{X'}_t$. $[\cdot]$ is the concatenation operation. GMH represents the graph multi-head attention block. $J$ denotes the number of GMH layers.

    In order to generate molecules in the reverse diffusion process, the trained score-based models $\boldsymbol{s}_\theta(\cdot,t)$ and $\boldsymbol{s}_\psi(\cdot,t)$ are used in Eq.~(\ref{equ:equ8}), then we have:
    \begin{equation}
    	\left\{\begin{array}{l}
    		\mathrm{d} \boldsymbol{X'}_t=[\mathbf{f}_{1, t}\left(\boldsymbol{X'}_t\right)-g_{1, t}^2 (\boldsymbol{s}_{\theta}\left(\boldsymbol{X'}_t, \boldsymbol{A'}_t,t\right)-\nabla_{\boldsymbol{X'}_t}\mathcal{E}_{val})] \mathrm{d} \bar{t}+g_{1, t} \mathrm{~d} \bar{\boldsymbol{\mathrm{w}}}_1\\
    		\mathrm{~d} \boldsymbol{A'}_t=[\mathbf{f}_{2,t}\left(\boldsymbol{A'}_t\right)-g_{2, t}^2 (\boldsymbol{s}_{\psi}\left(\boldsymbol{X'}_t, \boldsymbol{A'}_t,t\right)-\nabla_{\boldsymbol{A'}_t}\mathcal{E}_{val})]\mathrm{d} \bar{t}+g_{2,t} \mathrm{~d} \bar{\boldsymbol{\mathrm{w}}}_2
    	\end{array}\right.
    \label{equ:equ14}
    \end{equation}
    Finally, we use the Predictor-Corrector Sampler (PC sampler)~\cite{song2020score} to solve Eq.~(\ref{equ:equ14}) to generate molecules.

\subsection{Optimization for Multiple Molecular Properties}\label{sec:moo}
Our goal is to generate molecules with multiple desirable properties, we call it the \textit{optimization for multiple properties} (OMP) problem. Simply combing multiple properties into an indicator in a linear form and then optimizing the combined indicator does not work well here, as potential conflicts may exist among different properties
. Here, we transform it into a multi-objective optimization problem and solve it via Pareto Optimality~\cite{Mock2011}. 
 
Given a set of generated molecules 
$\{x_i,y^1_i,...,y^N_i\}_{i\in[K]}$ where $x_i$ a molecule with a series of properties $\{y^n\}_{n\in[N]}$ $y^n_i$. Define the loss function of the OMP module as follows:
    \begin{equation}
    	min_{\boldsymbol{\theta}}\mathcal{L}(\boldsymbol{\theta})=({\mathcal{L}_1(\boldsymbol{\theta})},{\mathcal{L}_2(\boldsymbol{\theta})},\cdots,{\mathcal{L}_N(\boldsymbol{\theta})})^T
    \end{equation}
    where $\mathcal{L}_n(\boldsymbol{\theta})$ is the loss of the $n$-th property of the generated molecule.


    \textbf{Pareto optimal solution.} Let $\boldsymbol{\theta}^*\in \Omega$, if there is no $\boldsymbol{\theta}\in \Omega$ that makes $\mathcal{L}_i(\boldsymbol{\theta})\le \mathcal{L}_i(\boldsymbol{\theta}^*)$, then $\boldsymbol{\theta}^*$ is a solution for the OMP problem. $\mathcal{L}_i(\boldsymbol{\theta}^*)$ is a Pareto optimal objective vector. $\mathcal{L}_i(\boldsymbol{\theta})\le \mathcal{L}_i(\boldsymbol{\theta}^*)$ means that $\boldsymbol{\theta}^*$ is the best solution without any potential improvement, which is called the Pareto optimal solution.

    \textbf{Pareto front.} The set of all Pareto optimal solutions $\boldsymbol{\theta}^*$ is called the Pareto set, which is denoted as $R_{pa}$. The image of the Pareto set is called the Pareto front ($P_f$) , where $P_f=\{\mathcal{L}(\boldsymbol{\theta})\}_{\boldsymbol{\theta} \in R_{pa}}$.


   Here we use 
   the Karush-Kuhn-Tucker (KKT) conditions~\cite{kuhn2013nonlinear}  to find the Pareto solutions. Formally, if there exist $\lambda_1,\ldots,\lambda_N\ge0$, where $\sum_{i=1}^N\lambda_i=1$ and $\sum_{i=1}^N\lambda_i\nabla \mathcal{L}_i(\boldsymbol{\theta}^i)=0$, then \{$\lambda_1,\ldots,\lambda_N$\} is a Pareto stationary point.

With the KKT conditions, the OMP problem is transformed to:
    \begin{equation}
    	\min _{\lambda_1, \ldots,\lambda_N}\left\{\left\|\sum_{i=1}^N \lambda_i \nabla \mathcal{L}_i\left(\boldsymbol{\theta}^i\right)\right\|_2^2 \bigg|\sum_{i=1}^N  \lambda_i=1, \lambda_i \geq 0 \quad \forall i\right\}
    \end{equation}
    When the above equation equals to zero, the corresponding solution satisfies the KKT conditions. Otherwise, a gradient descent direction is generated to further improve all the objectives~\cite{desideri2012multiple}.

    Specifically, as an example, in this paper we consider two properties: drug-likeness (QED) and PlogP, the optimization problem turns to the following loss function:
    \begin{equation}
    	\mathcal{L}_{OMP}=\min_{\lambda\in[0,1]}\left\|\lambda \nabla \mathcal{L}_Q\left(\boldsymbol{\theta}^Q\right)+(1-\lambda)\nabla \mathcal{L}_P\left(\boldsymbol{\theta}^P\right)\right\|_2^2
    	\label{equ:equ17}
    \end{equation}
   Eq.~(\ref{equ:equ17}) is a quadratic function of $\lambda$, which has the following solution:
    \begin{equation}
    	\hat{\lambda}=\left[\frac{\left(\nabla \mathcal{L}_P\left(\boldsymbol{\theta}^P\right)-\nabla{\mathcal{L}}_Q\left( 
    		\boldsymbol{\theta}^Q\right)\right)^{\top} \nabla {\mathcal{L}}_P\left(\boldsymbol{\theta}^P\right)}{\left\|\nabla {\mathcal{L}}_Q\left( \boldsymbol{\theta}^Q\right)-\nabla {\mathcal{L}}_P\left(\boldsymbol{\theta}^P\right)\right\|_2^2} \right]_{+, \underset{T}{1}}
    \end{equation}
    where $[\cdot]_{+, \underset{T}{1}}$ clips to $[0,1]$, as $[\lambda]_{+, \underset{T}{1}}=max(min(\lambda,1),0)$. 

    Though we consider only QED and PlogP here, our method is applicable to efficiently optimize more molecular properties simultaneously.

\section{Performance Evaluation}
Here we first describe the experimental setup, then present the experimental results, including the comparison with existing models and ablation studies. The implementation details and more experimental results are presented in the appendix.
	
\subsection{Experimental Setup}
\textbf{Datasets.} Two mainstream datasets QM9~\cite{ramakrishnan2014quantum} and ZINC250K~\cite{irwin2012zinc} are used in our experiments. QM9 includes 133,885 molecules grouped into 4 different types. Each molecule in this dataset contains up to 9 atoms. ZINC250K  has 249,455 drug-like molecules classified into 9 different types. Each molecule of this dataset contains up to 38 atoms. All molecules are kekulized by the chemical RDKit~\cite{landrum2016rdkit} with the hydrogen atoms being removed. The molecules have three types of edges, namely the single, double and triple bonds.

\textbf{Compared models.} Our model is compared with seven typical existing molecular generation models: a) the SOTA VAE-based method \textbf{JT-VAE}~\cite{jin2018junction}, which enhances the chemical validity of generated molecules by combining a tree-structured scaffold into a molecule through a graph message passing network; b) 4 flow-based methods: the normalizing flow-based model~\textbf{GraphNVP}~\cite{madhawa2019graphnvp} that generates valid molecular graphs with almost no duplicated ones; the autoregressive-flow-based model~\textbf{GraphAF}~\cite{shi2020graphaf} that generates molecules in a sequential way; \textbf{GraphDF}~\cite{luo2021graphdf} that uses invertible modulo shift transforms to map discrete latent variables to graph nodes and edges; the SOTA flow-based model~\textbf{MoFlow}~\cite{zang2020moflow} that learns the invertible mappings between molecular graphs and their latent representations; c) 2 diffusion-based methods: the SOTA model \textbf{GDSS}~\cite{jo2022score} and \textbf{DiGress}~\cite{vignac2022digress} --- a discretized diffusion model that generates graphs with categorical node and edge attributes. We consider no GAN and energy based models as they are not in the SOTA list. 
	
\textbf{Performance metrics.} Following previous works~\cite{shi2020graphaf,luo2021graphdf}, the quality of the 10,000 molecules generated by different models is evaluated with the following widely-used metrics: \textbf{Validity}, which refers to the fraction of chemically-correct molecules among all generated molecules. \textbf{Validity without correction (Val. w/o corr.)}~\cite{jo2022score}, which is used to compare the percentage of valid molecules without valency correction or edge resampling. \textbf{Uniqueness} refers to the percentage of distinct molecules among all valid molecules generated. \textbf{Fréchet ChemNet Distance (FCD)}~\cite{preuer2018frechet} measures the distance between the training data and the generated data by utilizing the activations of the penultimate layer of the ChemNet. \textbf{Quantitative Estimate of Drug-likeness (QED)}~\cite{bickerton2012quantifying} measures the drug-likeness of the generated molecules. \textbf{Penalized logP (PlogP)} is the log octanol-water partition coefficient (logP) score penalized by the ring size and synthetic accessibility.

\subsection{Experimental Results}
\subsubsection{Performance comparison}
We first compare some basic properties including Validity, Val. w/o corr., Uniqueness, Novelty and FCD between our method and the seven existing models. The results are presented in  Table~\ref{tab:1}. Note that some results of JT-VAE, GraphNVP, and DiGress are unavailable because they are unavailable in their original papers.  
We can see that our model outperforms the existing models in almost all these metrics on both QM9 and ZINC250K. Concretely, our model achieves 100\% validity and the highest validity w/o corr. on both datasets. It also ranks first in uniqueness on QM9, and has a high uniqueness of 99.88\% on ZINC250K. Following~\cite{vignac2022digress}, the novelty on QM9 is not considered since pursuing high novelty on this dataset will prevent models from capturing correct distributions. Our model reaches 100\% novelty on ZINC250K. Besides, our model significantly outperforms the existing models in FCD on both QM9 and ZINC250K, proving that the chemical space of our generated molecules is very close to the real data distribution. All the results above show that our model is more powerful in generating valid and drug-like molecules than the existing models.

We then compare the models in terms of two important molecular properties: QED and PlogP. We calculate the average QED and PlogP of the top-$k$ ($k$=1, 5, 10, 100, 1000) generated molecules as results, which are given in Tab.~\ref{tab:2}. As GraphAF cannot run on QM9, the results are unavailable.   
We can see that our model can generate molecules with higher QED and PlogP than the compared models, which shows that our model is able to generate molecules with desirable properties. 
      
      \begin{table}[h]
      	\centering
      	\caption{\textbf{Generation performance on QM9 and ZINC250k.}}
      	\label{tab:1}
      	\scalebox{0.7}{
      		\begin{tabular}{cccccccccc}
      			\hline
      			& \multicolumn{4}{c}{QM9}                                                                                                                                                                                                                  & \multicolumn{5}{c}{ZINC250k}                                                                                                                                                                                                                                                                                     \\ \cmidrule(r){2-5} \cmidrule(r){6-10}
      			Method                      & \begin{tabular}[c]{@{}c@{}}Validity\\ (\%)↑\end{tabular} & \begin{tabular}[c]{@{}c@{}}Val. w/o \\ corr.(\%)↑\end{tabular} & \begin{tabular}[c]{@{}c@{}}Uniqueness\\ (\%)↑\end{tabular} & \begin{tabular}[c]{@{}c@{}}FCD\\ ↓\end{tabular} & \begin{tabular}[c]{@{}c@{}}Validity\\ (\%)↑\end{tabular} & \begin{tabular}[c]{@{}c@{}}Val. w/o \\ corr.(\%)↑\end{tabular} & \begin{tabular}[c]{@{}c@{}}Uniqueness\\ (\%)↑\end{tabular} & \begin{tabular}[c]{@{}c@{}}Novelty\\ (\%)↑\end{tabular} & \begin{tabular}[c]{@{}c@{}}FCD\\ ↓\end{tabular}               \\ \hline
      			JT-VAE                      & n/a                                                      & n/a                                                            & n/a                                                        & n/a                                             & 100                                                      & n/a                                                            & \cellcolor[HTML]{D9D9D9}\textbf{100}                       & 100                                                     & 17.92                                                         \\
      			GraphNVP                    & 83.02                                                    & n/a                                                            & 99.23                                                      & 3.16                                            & 42.6                                                     & n/a                                                            & 94.8                                                       & 100                                                     & 39.62                                                         \\
      			GraphAF                     & 100                                                      & 67                                                             & 94.51                                                      & 5.27                                            & 100                                                      & 68                                                             & 99.10                                                      & 100                                                     & 16.29                                                         \\
      			GraphDF                     & 100                                                      & 82.67                                                          & 97.62                                                      & 10.82                                           & 100                                                      & 89.03                                                          & 99.16                                                      & 100                                                     & 34.20                                                         \\
      			MoFlow                      & 100                                                      & 96.17                                                          & 99.20                                                      & 4.47                                            & 100                                                      & 81.76                                                          & 99.99                                                      & 100                                                     & 20.93                                                         \\
GDSS                        & 100                                                      & 95.72                                                          & 97.82                                                      & 2.90                                            & 100                                                      & 97.01                                                          & 99.64                                                      & 100                                                     & 14.66                                                                                                  \\
      			DiGress & n/a                                                     & \cellcolor[HTML]{D9D9D9}\textbf{99.0}                                                            & 96.2                                                      & n/a                                             & n/a                                                      & n/a                                                            & n/a                                                        & n/a                                                     & n/a                                                           \\ \hline
      			\textbf{D2L-OMP~(ours)}         & \cellcolor[HTML]{D9D9D9}\textbf{100.00}                  & 98.60                         & \cellcolor[HTML]{D9D9D9}\textbf{99.80}                     & \cellcolor[HTML]{D9D9D9}\textbf{2.77}           & \cellcolor[HTML]{D9D9D9}\textbf{100.00}             & \cellcolor[HTML]{D9D9D9}\textbf{97.51}                         &
      			99.88                              & \cellcolor[HTML]{D9D9D9}\textbf{100.00}                 & \cellcolor[HTML]{D9D9D9}\textbf{14.05}\\ \hline
      	\end{tabular}}
      \end{table}
 
    \begin{table}[h]
    	\centering
    	\caption{\textbf{Comparison of chemical properties of the top-$k$ generated molecules.}}
    	\label{tab:2}
     \scalebox{0.7}{
\begin{tabular}{ccccccccccc}
\hline
\multirow{2}{*}{} & \multicolumn{5}{c}{QED ↑}                                                                 & \multicolumn{5}{c}{PlogP ↑}                                                               \\\cmidrule(r){2-6} \cmidrule(r){7-11}
                       Method & Top-1           & Top-5           & Top-10          & Top-100         & Top-1000        & Top-1           & Top-5           & Top-10          & Top-100         & Top-1000        \\ \cmidrule(r){1-6} \cmidrule(r){7-11}

GraphNVP                & 0.6344          & 0.6238          & 0.6204          & 0.5894          & 0.4166          & 2.8175          & 2.4048          & 2.1391          & 0.8324          & -1.6661         \\
GDSS                    & 0.6379          & 0.6269          & 0.6214          & 0.6121          & 0.4371          & 2.2792          & 2.2079          & 2.0266          & 1.3375          & -0.2180         \\
MoFlow                  & 0.6579          & 0.6544          & 0.6510          & 0.6188          & 0.5661          & 2.4654          & 1.6136          & 1.2339          & -0.0794         & -1.4578         \\ \hline
\textbf{D2L-OMP (ours)}     & \cellcolor[HTML]{D9D9D9}\textbf{0.6751} & \cellcolor[HTML]{D9D9D9}\textbf{0.6582} & \cellcolor[HTML]{D9D9D9}\textbf{0.6557} & \cellcolor[HTML]{D9D9D9}\textbf{0.6191} & \cellcolor[HTML]{D9D9D9}\textbf{0.5805} & \cellcolor[HTML]{D9D9D9}\textbf{2.9010} & \cellcolor[HTML]{D9D9D9}\textbf{2.5985} & \cellcolor[HTML]{D9D9D9}\textbf{2.4331} & \cellcolor[HTML]{D9D9D9}\textbf{1.6610} & \cellcolor[HTML]{D9D9D9}\textbf{0.1931} \\ \hline
\end{tabular}}

\centering
\scalebox{0.7}{
\begin{tabular}{ccccccccccc}
\hline
                         & \multicolumn{5}{c}{QED ↑}                                                                                                                                                                                         & \multicolumn{5}{c}{PlogP ↑}                                                                                                                                                                                       \\\cmidrule(r){2-6} \cmidrule(r){7-11} Method
\multirow{-2}{*}{} & Top-1                                   & Top-5                                   & Top-10                                  & Top-100                                 & Top-1000                                & Top-1                                   & Top-5                                   & Top-10                                  & Top-100                                 & Top-1000                                \\ \cmidrule(r){1-6} \cmidrule(r){7-11} 

GraphNVP                 & 0.8512                                  & 0.8391                                  & 0.8260                                  & 0.7423                                  & 0.5046                                  & 4.4533                                  & 3.7001                                  & 3.4829                                  & 2.2357                                  & -1.4607                                 \\
GraphAF                  & 0.9437                                  & 0.9261                                  & 0.9176                                  & 0.8593                                  & 0.5749                                  & 4.3435                                  & 3.6492                                  & 3.3885                                  & 2.5014                                  & -3.9328                                 \\
GDSS                     & 0.9449                                  & 0.9369                                  & 0.9337                                  & 0.9126                                  & 0.8512                                  & 3.3921                                  & 3.1951                                  & 3.0699                                  & 2.1932                                  & 0.2188                                  \\
MoFlow                   & 0.9261                                  & 0.9233                                  & 0.9150                                  & 0.8664                                  & 0.7839                                  & 3.2996                                  & 3.0081                                  & 2.8048                                  & 1.9469                                  & 0.3355                                  \\ \hline
\textbf{D2L-OMP~(ours)}      & \cellcolor[HTML]{D9D9D9}\textbf{0.9476} & \cellcolor[HTML]{D9D9D9}\textbf{0.9417} & \cellcolor[HTML]{D9D9D9}\textbf{0.9372} & \cellcolor[HTML]{D9D9D9}\textbf{0.9144} & \cellcolor[HTML]{D9D9D9}\textbf{0.8529} & \cellcolor[HTML]{D9D9D9}\textbf{4.6726} & \cellcolor[HTML]{D9D9D9}\textbf{3.7786} & \cellcolor[HTML]{D9D9D9}\textbf{3.5374} & \cellcolor[HTML]{D9D9D9}\textbf{2.6645} & \cellcolor[HTML]{D9D9D9}\textbf{1.2093} \\ \hline
\end{tabular}}
\end{table}

\subsubsection{Ablation studies}\label{sec:ablation}
First, we conduct ablation studies to check the effectiveness of 3 major modules: diffusion on fragments, energy-guidance function, and multi-property optimization. The results are presented in Table~\ref{tab:3}, from which we can see that by removing any of the three components, the performance of our method is degraded, indicating that they are all indispensable. 
Then, we examine the effect of our fragmentation method against five variants: using no fragments and no property optimization (baseline) and using BRICS~\cite{degen2008art},  FraGAT~\cite{zhang2021fragat}, CAFE-MPP~\cite{xie2023self}, and FREED~\cite{yang2021hit}. The results are in Table~\ref{tab:4}, from which We can see that the number of fragments gotten by our FREE method is an order of magnitude smaller than that gotten by BRICS~\cite{degen2008art},  FraGAT~\cite{zhang2021fragat} and CAFE-MPP~\cite{xie2023self}, which makes our model more efficient. More importantly, with fewer fragments, our model achieves better performance. This implies that the fragments generated by our method are more informative. 

The results of how some hyperparameters impact the model performance are presented in appendix.  

    
	\begin{table}[h]
		\centering
		\caption{\textbf{The effects of different variants of the D2L-OMP method.}}
		\label{tab:3}
  \scalebox{0.75}{
\begin{tabular}{ccccccc}
\hline
\multicolumn{1}{l}{}                & \multicolumn{3}{c}{QM9}                                                                                                                                                       & \multicolumn{3}{c}{ZINC250k}                                                                                                                                                  \\ \cmidrule(r){2-4} \cmidrule(r){5-7} 
Method                              & \begin{tabular}[c]{@{}c@{}}Val. w/o corr.\\ (\%)↑\end{tabular} & \begin{tabular}[c]{@{}c@{}}Uniqueness\\ (\%)↑\end{tabular} & \begin{tabular}[c]{@{}c@{}}FCD\\ ↓\end{tabular} & \begin{tabular}[c]{@{}c@{}}Val. w/o corr.\\ (\%)↑\end{tabular} & \begin{tabular}[c]{@{}c@{}}Uniqueness\\ (\%)↑\end{tabular} & \begin{tabular}[c]{@{}c@{}}FCD\\ ↓\end{tabular} \\ \cmidrule(r){1-4} \cmidrule(r){5-7} 
D2L-OMP w/o diffusion on fragments            & 96.06                                                          & 98.72                                                      & 3.41                                            & 95.34                                                          & 98.66                                                      & 15.57                                           \\
D2L-OMP w/o energy guidance function             & 97.00                                                          & 99.07                                                      & 3.15                                            & 96.61                                                          & 98.90                                                      & 22.34                                           \\
D2L-OMP w/o multi-property optimization & 93.15                                                          & 96.18                                                      & 4.46                                            & 95.83                                                          & 99.75                                                      & 17.62                                           \\ \hline
\textbf{D2L-OMP~(ours)}                                & \textbf{98.60}                                                          & \textbf{99.80}                                                      & \textbf{2.77}                                            & \textbf{97.51}                                                          &\textbf{99.88}                                                      & \textbf{14.05}                                           \\ \hline
\end{tabular}}
\end{table}

\begin{table}[h]
 \centering
    \caption{\textbf{Performance comparison of different fragmentation methods on QM9.}}    \label{tab:4}
\begin{tabular}{ccccc}
\hline
Method                           & number of fragments & Val. w/o corr.(\%)↑                    & Uniqueness(\%)↑                                               & FCD↓                                             \\ \hline
Baseline(QM9)                         & n/a                 & 92.23                                  & 65.05                                                         & 11.96                                            \\
BRICS                            & 81499               & 94.45                                  & 68.63                                                         & 11.18                                            \\
FraGAT                           & 3846                & \cellcolor[HTML]{FFFFFF}93.24          & 72.60                                                         & 10.83                                            \\
CAFE-MPP & 23537               & 93.92                                  & 69.97                                                         & \multicolumn{1}{l}{{\color[HTML]{24292F} 10.63}} \\
FREED                            & 66                  & 93.71                                  & 66.22                                                         & 12.18                                            \\ \hline
\textbf{FREE(ours)}                             & 3908                & \cellcolor[HTML]{D9D9D9}\textbf{95.36} & \cellcolor[HTML]{D9D9D9}\textbf{74.12} & \cellcolor[HTML]{D9D9D9}\textbf{10.41}           \\ \hline
\end{tabular}
\end{table}

\section{Conclusion}
In this paper, we propose a novel method to generate molecules with multiple desirable properties. The method is an expanded diffusion model with innovative designs: diffusing on two molecular structural levels and optimizing for multiple properties. It is the first work that generates molecules with multiple properties optimized simultaneously. The effectiveness and advantages of the proposed method are validated by extensive experiments. Future work will focus on exploring diffusion on more molecular structural levels and efficient optimization algorithms for more properties.  

	\bibliographystyle{unsrt}
	\bibliography{D2L-OMP.bib}
 \clearpage
{\Huge \textbf{Appendix}}	

Here, we present additional explanations on the proposed method and algorithm, more experimental results (parameter effects, property distribution and visualization of generated molecules), model implementation details, impacts and limitations of the proposed method.    
 
\section{Additional Explanations on the D2L-OMP Method}
Our D2L-OMP method consists of three major innovative modules: (a) the method of molecule \textit{fragmentation based on electronic effect} (\textbf{FREE} in short), (b) the \textit{generative model diffusing on two molecular structural levels} (\textbf{D2L} in short), and (c) the module of \textit{optimization for multiple molecular properties} (\textbf{OMP} in short). In what follows, we try to give more intuitive explanations on (or more vividly illustrate) the motivation or the mechanism of these modules above.

\subsection{The FREE Method}
\begin{figure}[!h]
\centering			\includegraphics[width=1\linewidth]{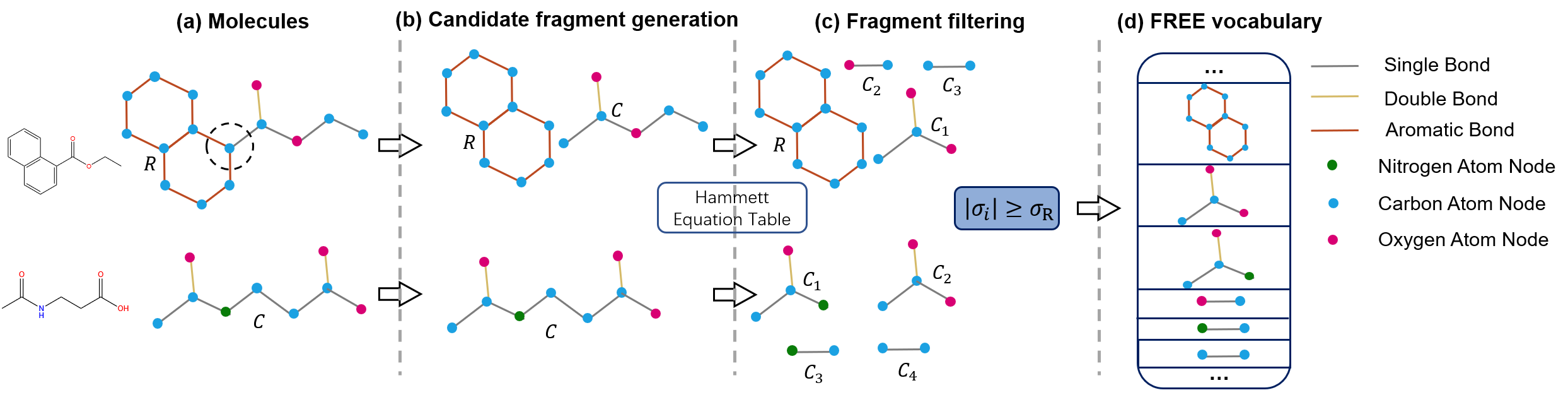}
\caption{The building process of the FREE Vocabulary.}\label{supfig:supfig1}
\end{figure}

Fig.~\ref{supfig:supfig1} illustrates the building process of the FREE vocabulary by two molecules, which are shown in Fig.~\ref{supfig:supfig1}(a). One (upper) has a double-ring substructure. Note that we treat any double-ring and multiple-ring substructure as a ring substructure. The other (down) has no ring substructure. 
Each training molecule is fragmented by the FREE method, introduced in Sec.3.2 and also restated below. 
 
Given a molecule, we fragment it as follows:
\begin{enumerate}
\item Split all ring substructures from their chain substructures, resulting in a set $R$=\{$R_i$\} of ring substructures and a set $C$=\{$C_i$\} of chain substructures, $R_i$ and $C_i$ are a ring substructure and a chain substructure respectively. 

\item We directly accept all elements in $R$ as fragments. 

\item For each element $C_i$ in $C$, first use the substituent with the highest constant in the Hammett table to search matching substructures in $C_i$, and accept the hits as fragments, meanwhile removing the matched substructures from $C_i$, and denoting the remaining as $C_i'$; Then, use the substituent with the second highest constant in the Hammett table to search matching substructures in $C_i'$, and save the hits as fragments, and the matched substructures are removed from $C_i'$. The above process goes iteratively on the remaining structure till no more match can be found. These operations are illustrated in Fig.~\ref{supfig:supfig1}(b)(c).

\item Using $\sigma_R$ as a threshold to filter out the fragments with constant $\left|\sigma\right|$ $<$ $\sigma_R$, the remaining fragments form the final fragment set. In our experiments, we set the  $R=$hydrogen atom, i.e., using the $\sigma_H$ as the threshold value. This is illustrated in Fig.~\ref{supfig:supfig1}(c).

\item  In such a way, we can get all fragments from the training molecules, which form a fragment vocabulary $V_{\boldsymbol{\mathcal{F}}}$, as shown in Fig.~\ref{supfig:supfig1}(d). In this paper, we call $V_{\boldsymbol{\mathcal{F}}}$ the \textit{FREE vocabulary}, and all fragments \textit{FREE fragments}.      
\end{enumerate}
		
\subsection{The D2L Model} 
\begin{figure}[h]
  		\centering
  		\includegraphics[width=1\linewidth]{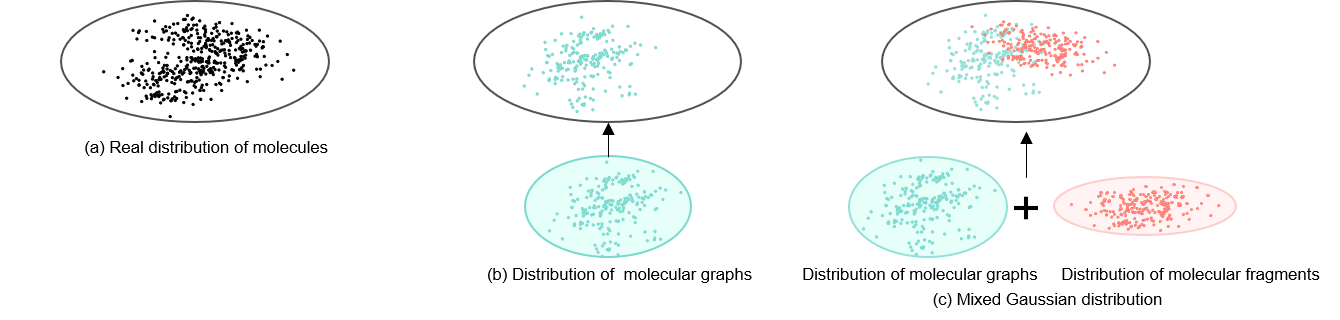}
  		\caption{Illustration of the motivation of two-level diffusion.}
  		\label{supfig:supfig2}
  	\end{figure}

Fig.~\ref{supfig:supfig2} is to illustrate the motivation of our D2L model, that is, why do we diffuse on two levels --- molecule graphs and molecular fragments? The rationale is like this: the structures of real molecules are complex and diverse, and diffusion only at the molecular level cannot capture the real molecular distribution (shown in Fig.~\ref{supfig:supfig2}(a)) well, as shown in Fig.~\ref{supfig:supfig2}(b). Therefore, we perform diffusion on two structural levels: molecules and molecular fragments, with which a mixed Gaussian distribution is obtained, and a better molecular distribution can be obtained, as shown in Fig.~\ref{supfig:supfig2}(c). 
   
\subsection{Energy Guidance Function}    
\begin{figure}[h]
  		\centering
  		\includegraphics[width=1\linewidth]{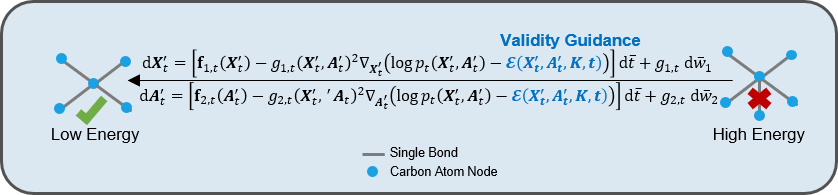}
  		\caption{Illustration of energy guidance function for validity optimization.}
  		\label{supfig:supfig3}
  	\end{figure}

Fig.~\ref{supfig:supfig3} illustrates how the energy guidance function guides the model to generate molecules of low energy (corresponding to high validity), while discarding molecules of high energy (corresponding to low validity), consequently optimizing the validity of the generated molecules. In our paper, we design the energy guidance function based on molecule validity. Certainly, we can design the energy guidance function for other properties.   
    
\subsection{Multiple-objective Optimization}   
\begin{figure}[h]
  		\centering
  		\includegraphics[width=1\linewidth]{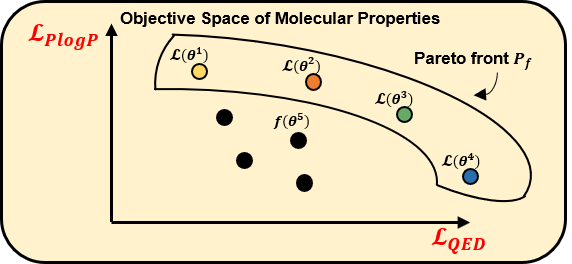}
  		\caption{Illustration of 2-property optimization by Pareto Optimal.}
  		\label{supfig:supfig4}
  	\end{figure}

Fig.~\ref{supfig:supfig4} illustrates optimizing two properties (QED and PlogP) simultaneously by Pareto optimal. As described in Sec.~3.5, let $\boldsymbol{\theta}^*\in \Omega$, if there is no $\boldsymbol{\theta}\in \Omega$ that makes $\mathcal{L}_i(\boldsymbol{\theta})\le \mathcal{L}_i(\boldsymbol{\theta}^*)$, then $\boldsymbol{\theta}^*$ is a solution for the OMP problem. $\mathcal{L}_i(\boldsymbol{\theta})\le \mathcal{L}_i(\boldsymbol{\theta}^*)$ means that $\boldsymbol{\theta}^*$ is the best solution without any potential improvement, which is called the Pareto optimal solution. The set of all Pareto optimal solutions $\boldsymbol{\theta}^*$ is called the Pareto set, which is denoted as $R_{pa}$. The image of the Pareto set is called the Pareto front ($P_f$) , where $P_f=\{\mathcal{L}(\boldsymbol{\theta})\}_{\boldsymbol{\theta} \in R_{pa}}$.

In Fig.~\ref{supfig:supfig4}, the molecules from $\mathcal{L}_i(\boldsymbol{\theta}^1)$ to $\mathcal{L}_i(\boldsymbol{\theta}^4)$ constitute the Pareto front, i.e., the optimal solution with regard to the properties QED and Plog, as no other molecules in the space have better QED and PlogP than those in the Pareto front. 

\section{Algorithm}
Alg.~\ref{alg:algorithm1} outlines the procedure of our method D2L-OMP, which 
consists of two parts, the diffusion process and the sampling process: the diffusion process aims to simultaneously diffuse at the two levels of molecules and fragments to obtain a mixed Gaussian distribution (Lines 2\textasciitilde5), and the sampling process aims to use the energy-guidance function and multiple -objective mechanisms to optimize the reverse process and sample new molecules from the obtained optimized distribution (Lines 6\textasciitilde15).
    	\begin{algorithm}[H]
    		\caption{The D2L-OMP algorithm.}
    		\label{alg:algorithm1}
    		\DontPrintSemicolon
    		\SetAlgoLined
    		\KwIn {A molecular graph $G$, a molecular fragment $\mathcal{F}$, denoising steps $N$, the score function $\boldsymbol{s}(\cdot,t)$}
    		\Begin{
    			$\Delta t=\frac{T}{N}$ \;
    			$G_0\sim p_{mol} $, $\mathcal{F}_0\sim p_{fra} $ \;
    			$G_T \gets G_0, \mathcal{F}_T \gets \mathcal{F}_0$ \;
    			$G'_T= G_T+\mathcal{F}_T$ \;
    			\For {$i = N$ to $1$}{
    				$t=i \Delta t$ \;
    				$G' \sim p_{0t}(G'_t|G_0)$  \tcp*[h]{sample from mixed Gaussian distribution} \;
    				$\mathcal{E} \gets \mathcal{E_V}(G',K_V,t)$\tcp*[h]{calculate the energy function} \;
    				$\mathcal{L}_{OMP}=OMP(\mathcal{L}_Q,\mathcal{L}_P)$ \tcp*[h]{calculate the OMP} \;
    				$F=\mathbf{f}(\boldsymbol{G'})-g(\boldsymbol{G'})^2(s(G')-\nabla_{G'}\mathcal{E})$\;
    				$\boldsymbol{z} \sim \mathcal{N}(\mathbf{0}, \boldsymbol{I})$ if $i>1$, else $\boldsymbol{z}=\mathbf{0}$\;
    				$G' \gets G'-\Delta t F+ g(t)\sqrt{\Delta t}\boldsymbol{z}$  \tcp*[h]{see Eq.(6) for details} \;
    				optimizer.step($\mathcal{L}_{OMP}(G')$)\;
    			}
    			\Return{$G'$}
    		}
    	\end{algorithm}

\section{More Experimental Results}

\subsection{Effects of Parameters}

\textbf{Effect of parameter $\mathrm{W}_f$.}  
Tab.~\ref{tab:5} shows how the weight of molecular fragments $\mathrm{W}_f$ impacts performance on QM9. We can see when $\mathrm{W}_f$ 
is set to 0.01, our D2L-OMP method achieves the best performance. 

\textbf{Effect of the substituent constant threshold $\sigma_R$.} 
$\sigma_R$ impacts the number of fragments generated by our FREE method. Here, we check the effect of $\sigma_R$ on the number of FREE fragments. We choose different values for $\sigma_R$, the results are presented in Tab.~\ref{tab:6}, from which we can see that $\sigma_R$ slightly impacts the number of FREE fragments. When the number of fragments is fixed, including as many kinds of fragments as possible will lead to better molecule generation results. In our paper, we set the  $R=$hydrogen atom, i.e., using $\sigma_H$ as the threshold value.

\textbf{Effect of parameter $C_V$.}
$C_V$ in Eq.~(4) is a hyper-parameter to control the optimization strength of the energy guidance function. Here check its effect on QM9, the results are presented in Tab.~\ref{tab:7}, which shows that our D2L-OMP achieves the best performance when $C_V$ is set to 0.1.

		\begin{table}[h]
			\centering
			\caption{\textbf{Effect of parameter $\mathrm{W}_f$ on QM9.}}
			\label{tab:5}
			\begin{tabular}{cccc}
				\hline
				$\mathrm{W}_f$ & \begin{tabular}[c]{@{}c@{}}Val. w/o corr.\\ (\%)↑\end{tabular} & \begin{tabular}[c]{@{}c@{}}Uniqueness\\ (\%)↑\end{tabular} & \begin{tabular}[c]{@{}c@{}}FCD\\ ↓\end{tabular} \\ \hline
				1.0   & 10.43                                                          & 96.26                                                      & 6.53                                            \\
				
				0.5   & 93.56                                                          & 61.89                                                      & 10.42                                           \\
				0.1   & 93.33                                                          & 73.59                                                      & 10.55                                           \\
				0.01  & \textbf{95.36}                                                          & \textbf{74.12}                                             & \textbf{10.41}                                  \\
				0.001 & 92.99                                                          & 74.10                                                      & 10.44                                           \\ \hline
			\end{tabular}
		\end{table}		
		\begin{table}[h]
			\centering
			\caption{\textbf{Results when using different values of threshold $\sigma_R$ of FREE.}}
			\label{tab:6}
			\begin{tabular}{cc}
				\hline
				$\sigma_R$ &  \#fragments \\ \hline
				R=hydrogen atom          & 3908                \\
				R=Fluoro atom            & 3907                \\
				R=Cyano                  & 3894                \\
				R=Nitro                  & 3890                \\
				R=Methoxy                & 3989                \\
				R=Methyl                 & 3902                \\
				R=Hydroxy                & 3898                \\ \hline
			\end{tabular}
		\end{table}

           \begin{table}[]
           \centering
           \caption{\textbf{Effect of parameter $C_V$ on QM9.}}
           \label{tab:7}
           \begin{tabular}{cccc}
           \hline
           $C_V$ & \begin{tabular}[c]{@{}c@{}}Val. w/o corr.\\ (\%)↑\end{tabular} & \begin{tabular}[c]{@{}c@{}}Uniqueness\\ (\%)↑\end{tabular} & \begin{tabular}[c]{@{}c@{}}FCD\\ ↓\end{tabular} \\ \hline
           1.0   & 99.64                                                          & 95.97                                                      & 3.98                                            \\
           0.5   & 99.40                                                          & 97.50                                                      & 3.43                                            \\
           0.1   & \textbf{98.60}                                                 & \textbf{99.80}                                             & \textbf{2.77}                                   \\
           0.01  & 98.55                                                          & 97.47                                                      & 3.35                                            \\
           0.001 & 98.57                                                          & 97.47                                                      & 3.36                                            \\ \hline
           \end{tabular}
           \end{table}

\subsection{Property Distribution of Generated Molecules}
\begin{figure}[h]
		\centering  
		\subfigure[QED]{
			\label{fig5.sub.1}
			\includegraphics[width=0.5\linewidth]{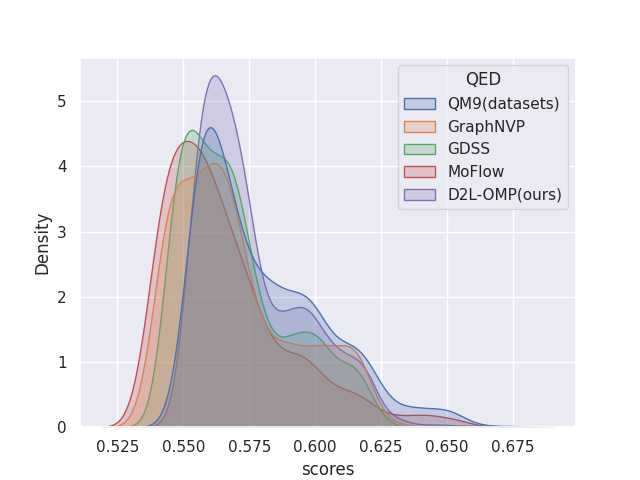}}\subfigure[PlogP]{\label{fig5.sub.2}
			\includegraphics[width=0.5\linewidth]{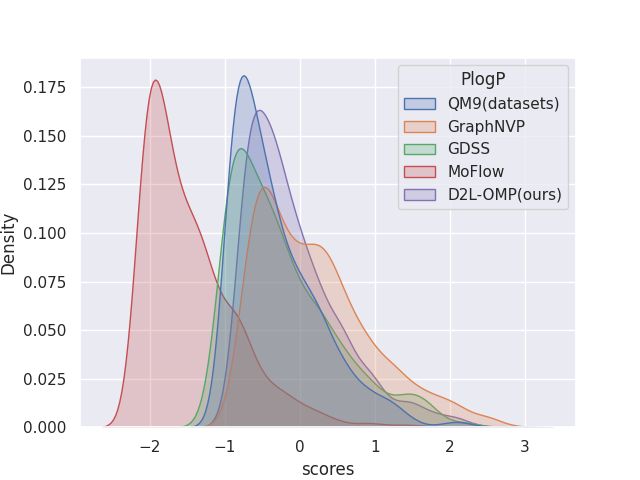}}
		\caption{Property Distributions of 10,000 molecules generated by different models on QM9.}
		\label{supfig:supfig5}
		\centering  
		\subfigure[QED]{
			\label{fig6.sub.1}
			\includegraphics[width=0.5\linewidth]{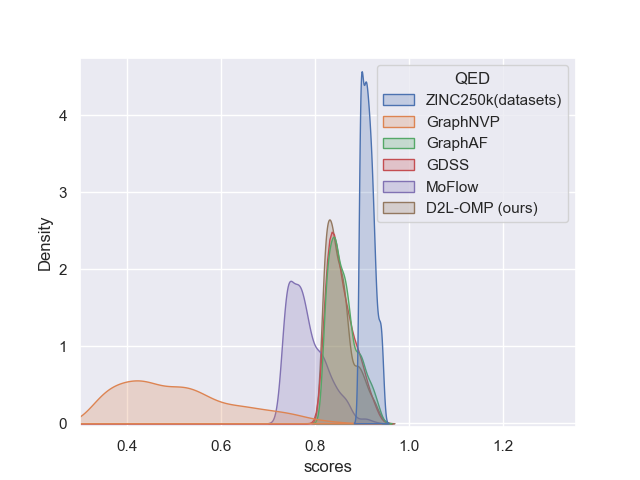}}\subfigure[PlogP]{\label{fig6.sub.2}
			\includegraphics[width=0.5\linewidth]{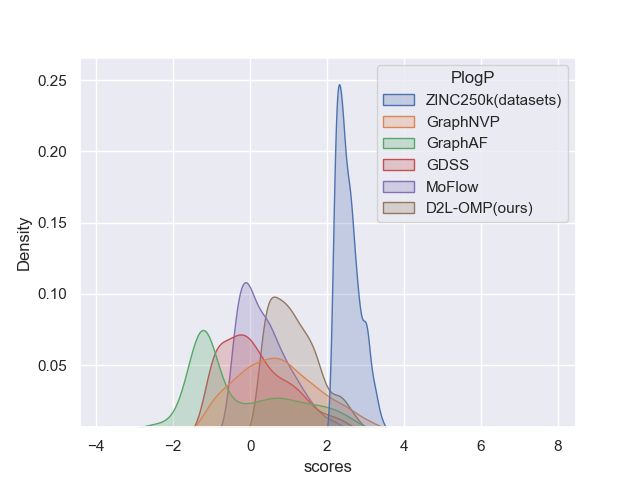}}
		\caption{Property Distributions of 10,000 molecules generated by different models on ZINC250k.}
		\label{supfig:supfig6}
	\end{figure}

Here, we compare property (QED and PlogP) distributions of 10,000 generated molecules by different models. 
The results are shown in Fig.~\ref{supfig:supfig5} and Fig.~\ref{supfig:supfig6}, from which we can see that the property distributions of molecules generated by our D2L-OPM method is closer to the property distributions of the datasets than the other models in most cases.

\subsection{Visualization of Generated Molecules}
\begin{figure}[h]
    	\centering
    	\includegraphics[width=0.8\linewidth]{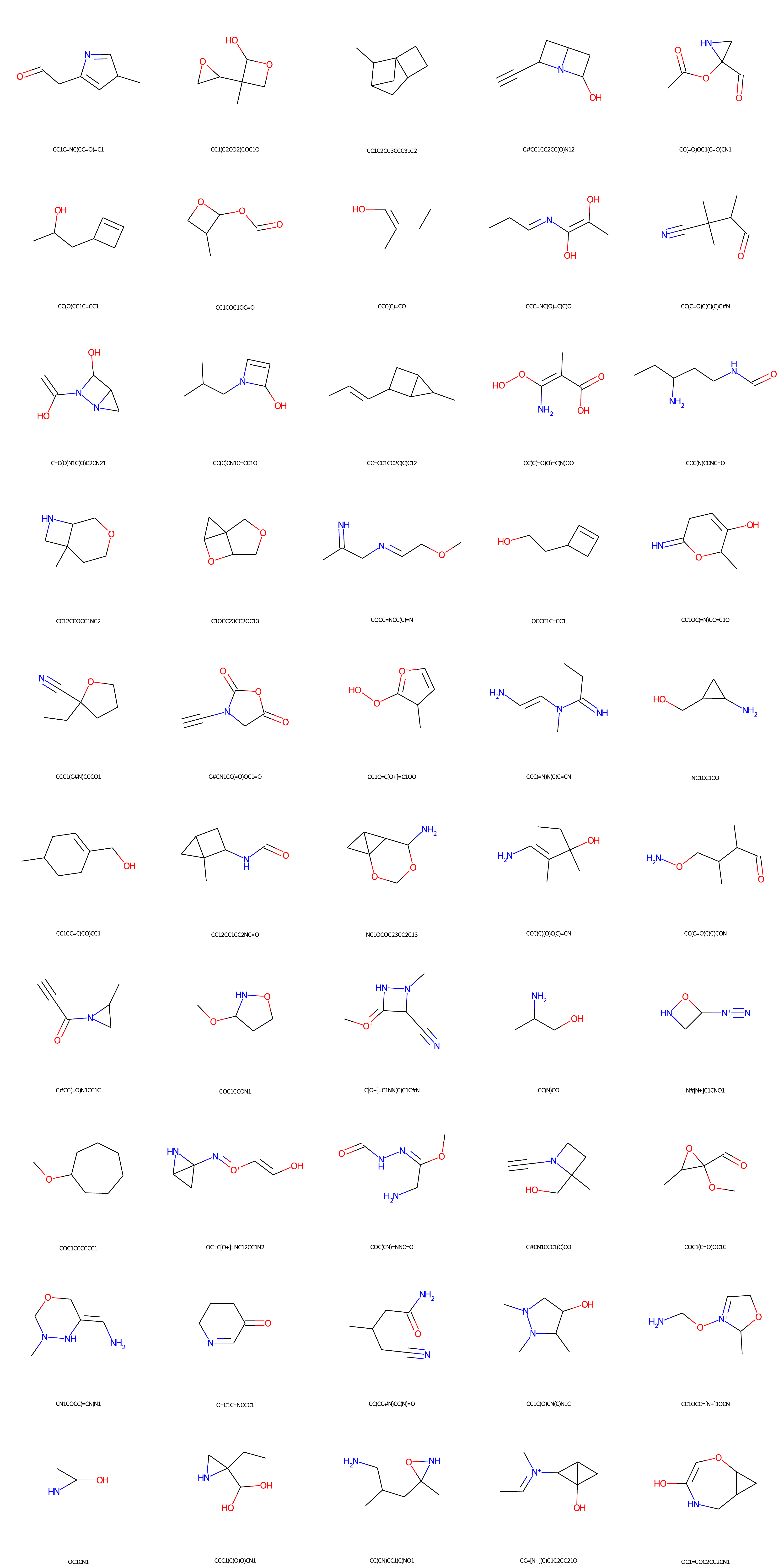}
    	\caption{Visualization of 50 randomly-selected molecules generated by our model on QM9.}
    	\label{supfig:supfig7}
    \end{figure}

\begin{figure}[h]
    	\centering
    	\includegraphics[width=0.8\linewidth]{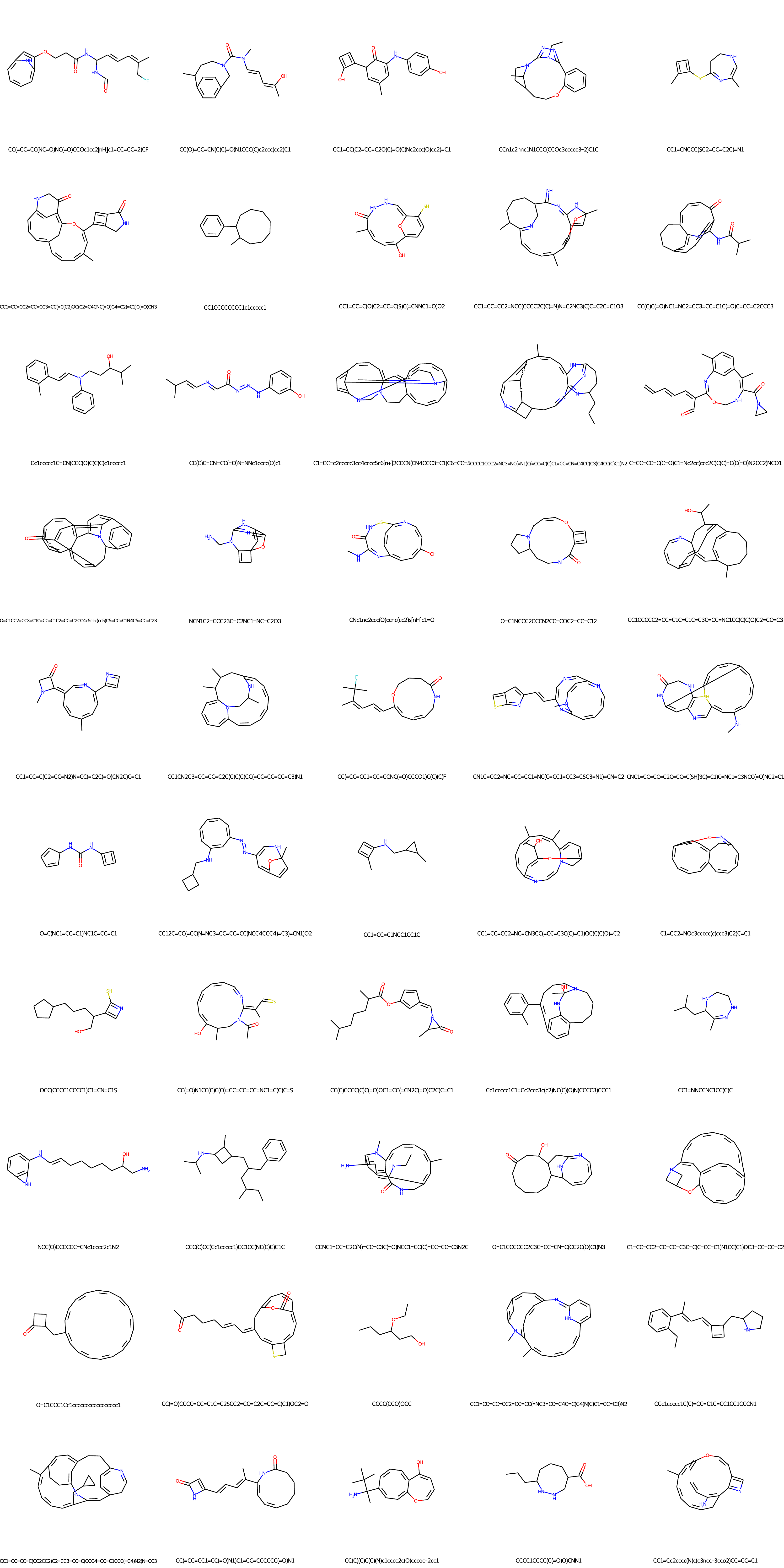}
    	\caption{Visualization of 50 randomly-selected molecules generated by our model on ZINC250k.}
    	\label{supfig:supfig8}
    \end{figure}

Fig.~\ref{supfig:supfig7} and Fig.~\ref{supfig:supfig8} show 50 randomly selected molecules generated by our method on QM9 and ZINC250k, respectively. 
    
\section{Model Implementation Details}
Our method is implemented under PyTorch. The score-based model $\boldsymbol{s}_\theta(\cdot,t)$ is implemented with 3 GCN layers, and the hidden dimension is set as 16. $\boldsymbol{s}_\psi(\cdot,t)$ is implemented with 3 GCN layers on QM9 and 6 GCN layers on ZINC250k. The hidden dimension of $\boldsymbol{s}_\psi(\cdot,t)$ is set to 16. Our model is trained with a batch size of 1024 on GeForce RTX 3090 GPU. We train the model on QM9 for 300 epochs and on ZINC250k for 500 epochs. We optimize our model by Adam with a fixed learning rate of 0.005.

\section{Impacts and Limitations}
\textbf{Impacts.} In this paper we propose a novel model for generating molecules with multiple desirable properties. Our method is an expanded diffusion model with multiple innovative designs, including diffusing on two molecular structural levels --- molecules and fragments, optimizing multiple properties simultaneously, and fragmenting molecules based on electronic effect. Extensive experiments show that our method can generate molecules with better properties, including validity, uniqueness, novelty, FCD, QED and PlogP, than those generated by existing models. Our work pioneers a new way to molecule generation, thus is of considerable significance to drug design and discovery.

\textbf{Limitations.} Our method is an expanded diffusion model, integrating with multiple-objective optimization. Both the diffusion model and multiple-objective optimization are relatively time-consuming. So in the future, we will explore efficient diffusion models and multiple-objective optimization algorithms. Furthermore, here we optimize three properties simultaneously. But molecules have many properties. So as future work, we will try to optimize more properties simultaneously. 
	

\end{document}